\documentclass[aps,prl,twocolumn,superscriptaddress,showpacs,english]{revtex4}

\usepackage{graphicx}
\usepackage{epstopdf}
\usepackage{dcolumn}
\usepackage{bm}
\usepackage{color}
\usepackage{multirow}

\definecolor{darkgreen}{rgb}{0.0,0.5,0.0}

\usepackage[linktocpage=true,
  colorlinks=true, 
  pdfborder={0 0 0},
  linkcolor=blue,
  citecolor=blue,
  filecolor=yellow,
  urlcolor=darkgreen,
  bookmarks,
  pdfauthor={},
]{hyperref}

\begin{document}
\newcommand{\kvec}{\mbox{{\scriptsize {\bf k}}}}
\def\eq#1{(\ref{#1})}
\def\fig#1{\hspace{1mm}Fig. \ref{#1}}
\def\tab#1{\hspace{1mm}\ref{#1}}
{\color{black}
\title{High-temperature study of superconducting hydrogen and deuterium sulfide}
\author{A.P. Durajski}\email{adurajski@wip.pcz.pl} 
\affiliation{Institute of Physics, Cz{\c{e}}stochowa University of Technology, Ave. Armii Krajowej 19, 42-200 Cz{\c{e}}stochowa, Poland}
\author{R. Szcz{\c{e}}{\'s}niak}
\affiliation{Institute of Physics, Cz{\c{e}}stochowa University of Technology, Ave. Armii Krajowej 19, 42-200 Cz{\c{e}}stochowa, Poland}
\affiliation{Institute of Physics, Jan D{\l}ugosz University, Ave. Armii Krajowej 13/15, 42-200 Cz{\c{e}}stochowa, Poland}
\author{L. Pietronero}
\affiliation{Sapienza, Universit{\`a} di Roma, Dip. Fisica, P. le A. Moro 2, 00185 Roma, Italy}
\affiliation{Institute of Complex Systems, CNR, Via dei Taurini 19 Roma, Italy}
\affiliation{London Institute for Mathematical Sciences, South Street 22, Mayfair London, United Kingdom}
\date{\today} 

%
\begin{abstract}
%
Hydrogen-rich compounds are extensively explored as candidates for a high-temperature superconductors. Currently, the measured critical temperature of $203$ K in hydrogen sulfide (H$_3$S) is among the highest over all-known superconductors.
In present paper, using the strong-coupling Eliashberg theory of superconductivity, we compared in detail the thermodynamic properties of two samples containing different hydrogen isotopes H$_3$S and D$_3$S at $150$ GPa.
Our research indicates that it is possible to reproduce the measured values of critical temperature $203$ K and $147$ K for H$_3$S and D$_3$S by using a Coulomb pseudopotential of $0.123$ and $0.131$, respectively. 
However, we also discuss a scenario in which the isotope effect is independent of pressure and the Coulomb pseudopotential for D$_3$S is smaller than for H$_3$S.
For both scenarios, the energy gap, specific heat, thermodynamic critical field and related dimensionless ratios are calculated and compared with other conventional superconductors.
We shown that the existence of the strong-coupling and retardation effects in the  systems analysed result in significant differences between values obtained within the framework of the Eliashberg formalism and the prediction of the Bardeen-Cooper-Schrieffer theory.
\end{abstract}
\maketitle

%
\section{I. Introduction}
%
Hydrogen sulfide at high pressure ($p$) is believed to be a phonon-mediated superconductor characterized by the highest superconducting transition temperature ($T_C$) ever measured. 
The recent electrical, resistance, magnetic susceptibility and Raman spectra measurements performed using a diamond anvil cell by Drozdov \textit{et al.} showed that H$_2$S transforms into a metal and then into a superconductor around $90$--$100$ GPa \cite{Drozdov2015A, Drozdov2014A}.
In a sample prepared at $p{<}150$ GPa and at low temperature of $100$--$150$ K, $T_C$ increases with pressure from $31$ K at $115$ GPa to $60$ K at $145$ GPa.
Then, as pressure increases above $150$ GPa a drastic increase of $T_C$ was observed and finally $T_C$ reaches $150$ K at $200$ GPa \cite{Drozdov2015A}.
The above experiment was directly inspired by previous theoretical calculations conducted within the density functional perturbation theory by Li \textit{et al.} \cite{Li2014A}. It should be noted that the measured results for $p{<}150$ GPa are in general agreement with the theoretical predictions mentioned.
Additional experimental measurements on a sample prepared at $p{>}150$ GPa and at high temperature of $220$--$300$ K found superconductivity with $T_C$ as high as $203$ K, breaking all the records thus far \cite{Drozdov2015A}.

These differences between samples prepared at low/high temperature and pressure are probably associated with the dissociation of solid hydrogen sulfide. Theoretical prediction clearly suggests the pressure-induced decomposition of H$_2$S to H$_3$S and elemental sulfur according to the following energetically allowed scheme: ${\rm 3H_{2}S\rightarrow 2H_{3}S+S}$ \cite{Duan2014A, Duan2014B, BernsteinH2S}. It means that H$_3$S, above $50$ GPa remains the more stable stoichiometry than H$_2$S, which suggests that the highest critical temperatures observed experimentally come from H$_3$S compound \cite{Errea, BernsteinH2S}.

This discovery has stimulated significant interest in studying the underlying superconducting mechanism \cite{BernsteinH2S, HirschH2S, Bianconi, Papaconstantopoulos, Banacky, Quan} and searching for novel methods to decrease the external hydrostatic pressure and increase the critical temperature of dense hydrogen sulfide \cite{Heil}.
The systematic theoretical study performed in paper \cite{DurajskiH2S} showed that the maximum value of the critical temperature in the family of the H$_n$S-type compounds ($n{\geq}2$) can even be equal to ${\sim}290$ K. According to the results of this work, hydrogen sulfide is a candidate for a high temperature superconductor with a critical temperature as high as the room temperature.

Motivated by the recent experimental and theoretical progress in this area, we have carried out calculations to explore in detail the thermodynamic properties of superconducting hydrogen sulfide H$_3$S and deuterium sulfide D$_3$S at high pressure.
In particular, we chose a pressure equal to ${\sim}150$ GPa due to the fact that the experimental results exist for both systems. Moreover, $T_C$ for H$_3$S is close to the maximum measured value.
The very large values of electron-phonon coupling strength $[\lambda]_{\rm H_3S}=2.067$ and $[\lambda]_{\rm D_3S}=2.065$ calculated using \textit{ab initio} methods \cite{AkashiH2S}, caused that our investigations were conducted within the framework of the strong-coupling Eliashberg theory, which allows us to describe the thermodynamic properties of conventional superconductors with experimental accuracy.

The paper is organized as follows. Section II contains a short outline of the strong-coupling Eliashberg formalism. In Section III, we discuss and compare the thermodynamic properties of superconducting H$_3$S and D$_3$S systems at $150$ GPa. Section IV summarizes the obtained results.

%
\section{II. Eliashberg formalism}
%
The finite temperature Eliashberg equations for the superconducting energy gap $\Delta\left(\omega\right)\equiv \phi\left(\omega\right)/Z\left(\omega\right)$ and the mass renormalization function $Z\left(\omega\right)$ 
written in a mixed representation (formulated of both the real and imaginary frequency axis) are given by \cite{Marsiglio1988A}:
\begin{eqnarray}
\label{r1}
\phi\left(\omega\right)&=&
\frac{\pi}{\beta}\sum_{m=-M}^{M}\frac{\left[\lambda\left(\omega-i\omega_{m}\right)-\mu^{\star}\theta\left(\omega_{c}-|\omega_{m}|\right)\right]}
{\sqrt{\omega_m^2Z^{2}_{m}+\phi^{2}_{m}}}\phi_{m}\\ \nonumber
                              &+& i\pi\int_{0}^{+\infty}d\omega^{'}\alpha^{2}F\left(\omega^{'}\right)
                                 \Big[\left[N\left(\omega^{'}\right)+f\left(\omega^{'}-\omega\right)\right]\\ \nonumber
                              &\times&K\left(\omega,-\omega^{'}\right)\phi\left(\omega-\omega^{'}\right)\Big]\\ \nonumber
                              &+& i\pi\int_{0}^{+\infty}d\omega^{'}\alpha^{2}F\left(\omega^{'}\right)
                                 \Big[\left[N\left(\omega^{'}\right)+f\left(\omega^{'}+\omega\right)\right]\\ \nonumber
                              &\times&K\left(\omega,\omega^{'}\right)\phi\left(\omega+\omega^{'}\right)\Big]
\end{eqnarray}
and
\begin{eqnarray}
\label{r2}
Z\left(\omega\right)&=&
                                  1+\frac{i\pi}{\omega\beta}\sum_{m=-M}^{M}
                                  \frac{\lambda\left(\omega-i\omega_{m}\right)\omega_{m}}{\sqrt{\omega_m^2Z^{2}_{m}+\phi^{2}_{m}}}Z_{m}\\ \nonumber
                              &+&\frac{i\pi}{\omega}\int_{0}^{+\infty}d\omega^{'}\alpha^{2}F\left(\omega^{'}\right)
                                  \Big[\left[N\left(\omega^{'}\right)+f\left(\omega^{'}-\omega\right)\right]\\ \nonumber
                              &\times&K\left(\omega,-\omega^{'}\right)\left(\omega-\omega^{'}\right)Z\left(\omega-\omega^{'}\right)\Big]\\ \nonumber
                              &+&\frac{i\pi}{\omega}\int_{0}^{+\infty}d\omega^{'}\alpha^{2}F\left(\omega^{'}\right)
                                  \Big[\left[N\left(\omega^{'}\right)+f\left(\omega^{'}+\omega\right)\right]\\ \nonumber
                              &\times&K\left(\omega,\omega^{'}\right)\left(\omega+\omega^{'}\right)Z\left(\omega+\omega^{'}\right)\Big], 
\end{eqnarray}
where
\begin{equation}
\label{r3}
K\left(\omega,\omega^{'}\right)\equiv
\frac{1}{\sqrt{\left(\omega+\omega^{'}\right)^{2}Z^{2}\left(\omega+\omega^{'}\right)-\phi^{2}\left(\omega+\omega^{'}\right)}}.
\end{equation}
Here, $\beta=1/k_BT$, symbol $\mu^{\star}$ means the screened Coulomb repulsion, $f\left(\omega\right)$ and $N\left(\omega\right)$ are the Fermi-Dirac and Bose-Einstein distribution functions. 
Symbols $\theta$ and $\omega_c$ denote the Heaviside function and the cut-off frequency, respectively. For $\omega_c$ we chosen ten times the maximum phonon frequency: $\omega_c=10\Omega_{\rm max}$.
The imaginary axis functions $\phi_{n}\equiv\phi\left(i\omega_{n}\right)$ and $Z_{n}\equiv Z\left(i\omega_{n}\right)$ are given by \cite{Eliashberg1960A}: 
\begin{equation}
\label{r4}
\phi_{n}=\frac{\pi}{\beta}\sum_{m=-M}^{M}
\frac{\lambda\left(i\omega_{n}-i\omega_{m}\right)-\mu^{\star}\left(\omega_{m}\right)}
{\sqrt{\omega_m^2Z^{2}_{m}+\phi^{2}_{m}}}\phi_{m}
\end{equation}
and
\begin{equation}
\label{r5}
Z_{n}=1+\frac{1}{\omega_{n}}\frac{\pi}{\beta}\sum_{m=-M}^{M}
\frac{\lambda\left(i\omega_{n}-i\omega_{m}\right)}{\sqrt{\omega_m^2Z^{2}_{m}+\phi^{2}_{m}}}
\omega_{m}Z_{m}.
\end{equation}
The Matsubara frequency is defined as $\omega_{n}\equiv\left(\pi/\beta\right)\left(2n-1\right)$ where $n=0,\pm 1,\pm 2,\dots,\pm M$, and $M=1100$. 
The pairing kernel for the electron-phonon interaction is given by:
\begin{equation}
\label{r6}
\lambda\left(z\right)\equiv 2\int_0^{\Omega_{\rm{max}}}d\omega\frac{\omega}{\omega ^2-z^{2}}\alpha^{2}F\left(\omega\right),
\end{equation}
\begin{figure}[!b]
\includegraphics[width=\columnwidth]{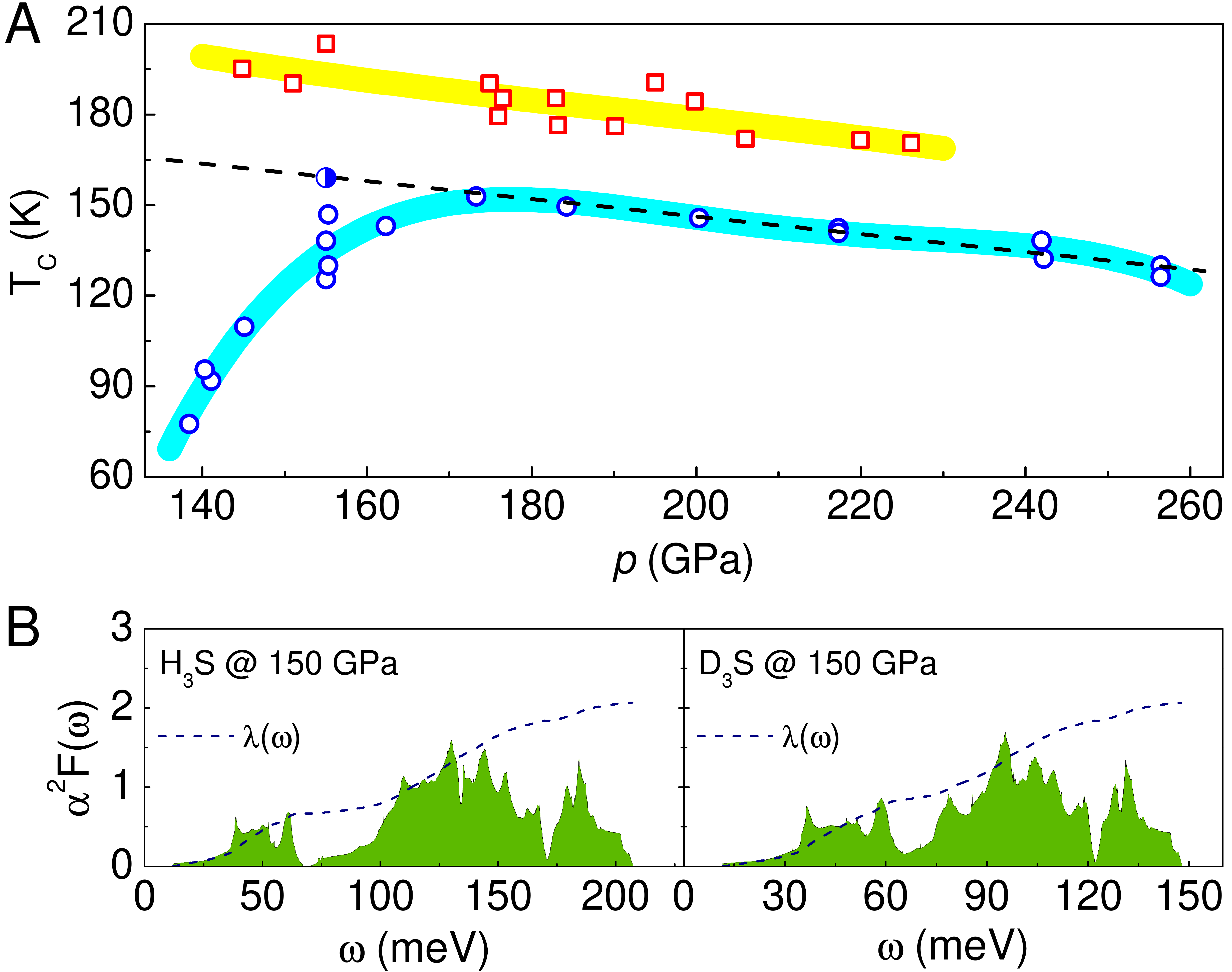}
\caption{(A) The pressure dependence of critical temperature for H$_3$S (open squares) and D$_3$S (open circles) \cite{Drozdov2015A}. The dashed line and half-filled circle represent our prediction related to the constant value of isotope effect. (B) The Eliashberg spectral functions $\alpha^2F(\omega )$ and the electron-phonon integrals $\lambda(\omega )$ of H$_3$S and D$_3$S at $150$ GPa \cite{AkashiH2S}.}
\label{f1}
\end{figure}
The Eliashberg spectral function $\alpha^{2}F(\omega)$ is expressed as a sum over the contributions from scattering processes which connect electrons through phonons on the Fermi surface.
Our investigations are based on the spectral function determined by using the \textit{ab initio} plane-wave pseudopotential calculations in paper \cite{AkashiH2S}:
\begin{eqnarray}
\alpha^{2}F(\omega)&=&\frac{1}{N(0)}
\sum_{{\nu {\bf q}nn'{\bf k}}}|g^{n{\bf k}+{\bf q},n'{\bf k}}_{\nu{\bf q}}|^{2}
\delta(\xi_{n{\bf k}+{\bf q}}) \\\nonumber
&\times&\delta(\xi_{n'{\bf k}})
\delta(\omega-\omega_{\nu{\bf q}}),
\label{eq:a2F}
\end{eqnarray}
where $N(0)$, $g^{n{\bf k}+{\bf q},n'\!{\bf k}}_{\nu{\bf q}}$ and $\omega_{\nu{\bf q}}$ denote the density of states at the Fermi energy, the electron-phonon matrix element and the phonon frequency, respectively.

The shape of the Eliashberg functions for H$_3$S and D$_3$S is presented in \fig{f1} (B).
Moreover, the dashed line is the integration of the electron-phonon coupling strength as a function of the phonon frequency:
\begin{equation}
\label{lambda}
\lambda(\omega)=2\int_0^{\Omega_{\rm{max}}}\frac{\alpha^2F(\omega)}{\omega}d\omega.
\end{equation}
The main part of \fig{f1} shows the experimental values of critical temperature for H$_3$S (open squares) and D$_3$S (open circles) as a function of pressure \cite{Drozdov2015A}.
Both from the experimental and theoretical point of view, especially interesting seems to be non-conventional behaviour of the isotope effect associated with low values of critical temperature for D$_3$S at low pressures.
This can result from inaccuracy of measurement, instability of the system, or it can be a new effect, not observed  until now.
Therefore, we discuss two scenarios. The first assuming the correctness of the experimental results, $T_C=203$ K for H$_3$S and $T_C=147$ K for D$_3$S \cite{Drozdov2015A}, and the second assuming invariance of the isotope effect, $T_C=203$ K for H$_3$S and $T_C=159$ K for D$_3$S (half-filled circle in \fig{f1}~(A)).

In order to determine the all significant thermodynamic properties of H$_3$S and D$_3$S systems, we use a standard Matsubara technique to solve the Eliashberg equations. In particular, we conducted numerical calculations based on a self-consistent iteration methods \cite{szczesniak2006A}, which were used successfully in our previous papers \cite{Durajski2015, GaH3, trojwodorki}.

%
\section{III. The Results and Discussion}
%
The isotope effect of superconducting critical temperature is best described in terms of the isotope effect coefficient ($\alpha$)
defined by the relation:
\begin{equation}
\label{alpha}
\alpha=-\frac{ {\rm ln}[T_{C}]_{{\rm D}_{3}{\rm S}}-{\rm ln}[T_{C}]_{{\rm H}_{3}{\rm S}}}{{\rm ln}[M]_{\rm D}-{\rm ln}[M]_{\rm H}},
\end{equation}
where $[M]_{\rm H}$ and $[M]_{\rm D}$ are the atomic mass of hydrogen and deuterium, respectively. In the case of experimental results for hydrogen and deuterium sulfide at $p=150$ GPa we have $\alpha=0.47$. This value is very close to the BCS value ($[\alpha]_{\rm BCS}\approx 0.5$). However, if we take into account $T_C=159$ K for D$_3$S, then $\alpha=0.35$, which is in a general agreement with the results obtained for a higher pressures.

To study the thermodynamic properties on the quantitatively level in the framework of the Eliashberg formalism we must first determine the critical value of the Coulomb pseudopotential $\mu^{\star}_C$.
For this purpose, in the Eliashberg equations we assumed that $T=T_C$ and then we increased the value of the parameter $\mu^{\star}$ until we reached the equality $\Delta_{m=1}=0$ at $\mu^{\star}=\mu^{\star}_C$. The full dependence of $\Delta_{m=1}$($\mu^{\star}$) are presented in \fig{f2}. 
On this basis we can notice that for the investigated systems, Coulomb pseudopotentials, correlated with experimental results take a typical critical values, $\mu^{\star}_C=0.123$ for H$_3$S and $\mu^{\star}_C=0.131$ for D$_3$S.
However, untypical is the fact that $[\mu^{\star}_C]_{\rm H_3S}<[\mu^{\star}_C]_{\rm D_3S}$, because according to the Morel-Anderson formula \cite{Morel1962A}:
\begin{equation}
\label{alpha}
\mu^{\star}=\frac{\mu}{1+ \mu {\rm ln}\left(\frac{\omega_{el}}{\omega_{ln}} \right) }
\simeq \frac{1}{{\rm ln}\left(\frac{\omega_{el}}{\omega_{ln}} \right) },
\end{equation}
$\mu^{\star}$ is proportional to the average phonon frequency ($\omega_{ln}$) which relates to the dynamics of the superconducting state:
$\omega_{{ln}}\equiv \exp\left[\frac{2}{\lambda}
\int^{+\infty}_{0}d\Omega\frac{\alpha^{2}F\left(\Omega\right)}
{\Omega}\ln\left(\Omega\right)\right]$. Due to the atom mass the average phonon frequency is larger for the system containing hydrogen than for the one containing deuterium, so we should observe that $[\mu^{\star}_C]_{\rm H_3S}>[\mu^{\star}_C]_{\rm D_3S}$. This situation is possible when we take into account $T_C=159$ K for D$_3$S. Then, $[\mu^{\star}_C]_{\rm D_3S}=0.088$. Such a small value of $\mu^{\star}_C$ is physically possible in the hydrogen rich materials, as was confirmed by the recent research of platinum hydride \cite{DominPtH}.

\begin{figure}[!h]
\includegraphics[width=\columnwidth]{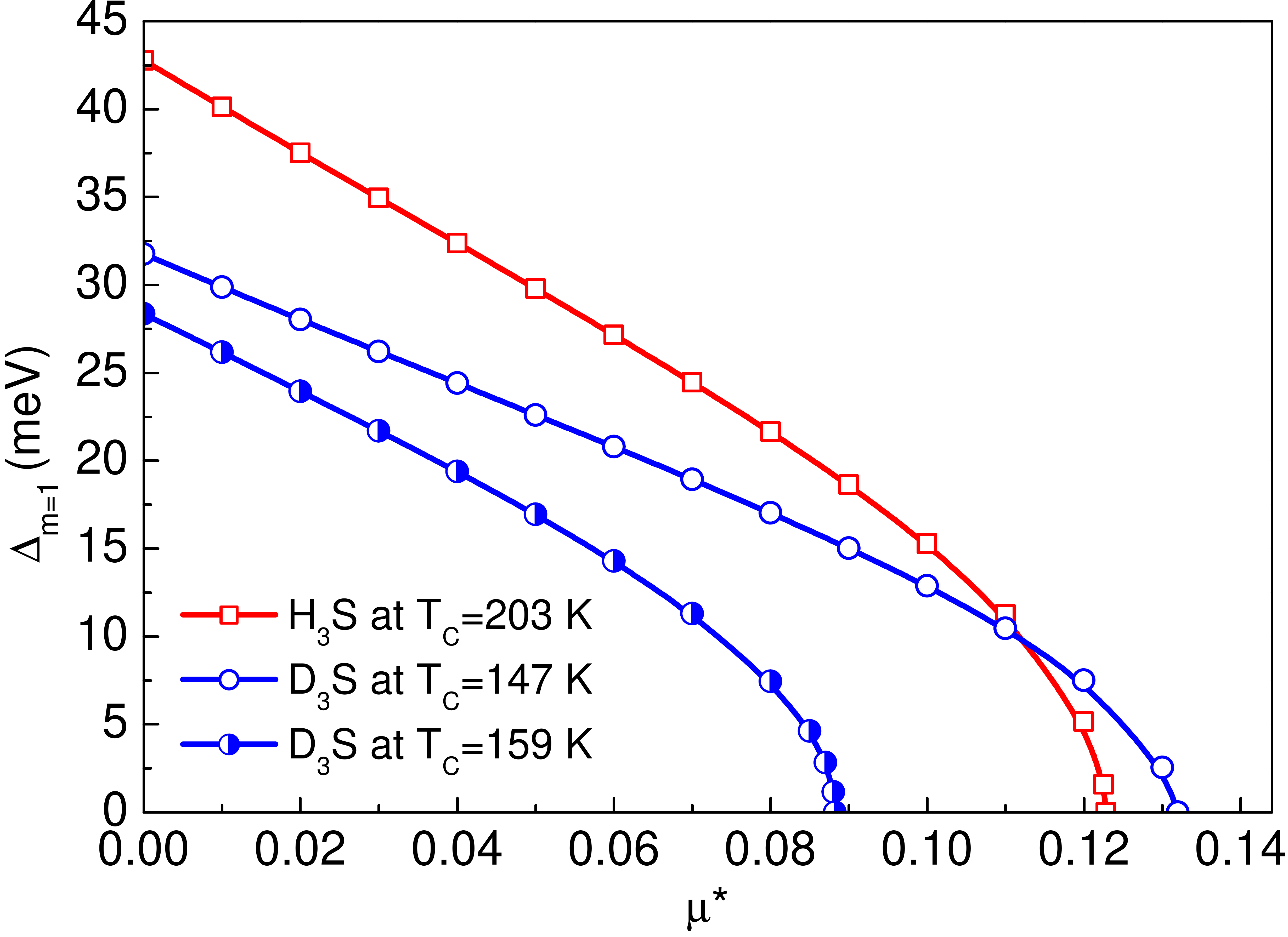}
\caption{The full dependence of the maximum value of the order parameter on the Coulomb
pseudopotential.}
\label{f2}
\end{figure}

Let us notice, that for the above values of $\mu^{\star}_C$, the critical temperatures observed in the experiment cannot be correctly reproduced using the analytical formulas for $T_C$. 
In particular, the Allen-Dynes expression \cite{Allen1975A} predicts that $[T_C(\mu^{\star}_C)]_{\rm H_3S}=176$~K, $[T_C(\mu^{\star}_C)]^{\rm (1)}_{\rm D_3S}=130$~K and $[T_C(\mu^{\star}_C)]^{\rm (2)}_{\rm D_3S}=147$ K; while the McMillan formula \cite{McMillan1968A} gives even more inaccurate results: $[T_C(\mu^{\star}_C)]_{\rm H_3S}=148$~K, $[T_C(\mu^{\star}_C)]^{\rm (1)}_{\rm D_3S}=110$~K and $[T_C(\mu^{\star}_C)]^{\rm (2)}_{\rm D_3S}=121$~K.
It should be noted that in order to improve readability of this paper we use in the superscript the abbreviation $(1)$ in the context of the first scenario ($T_C=147$ K for D$_3$S) and abbreviation $(2)$ in the context of the second scenario ($T_C=159$ K for D$_3$S) discussed in this paper.

\begin{figure}[!b]
\includegraphics[width=\columnwidth]{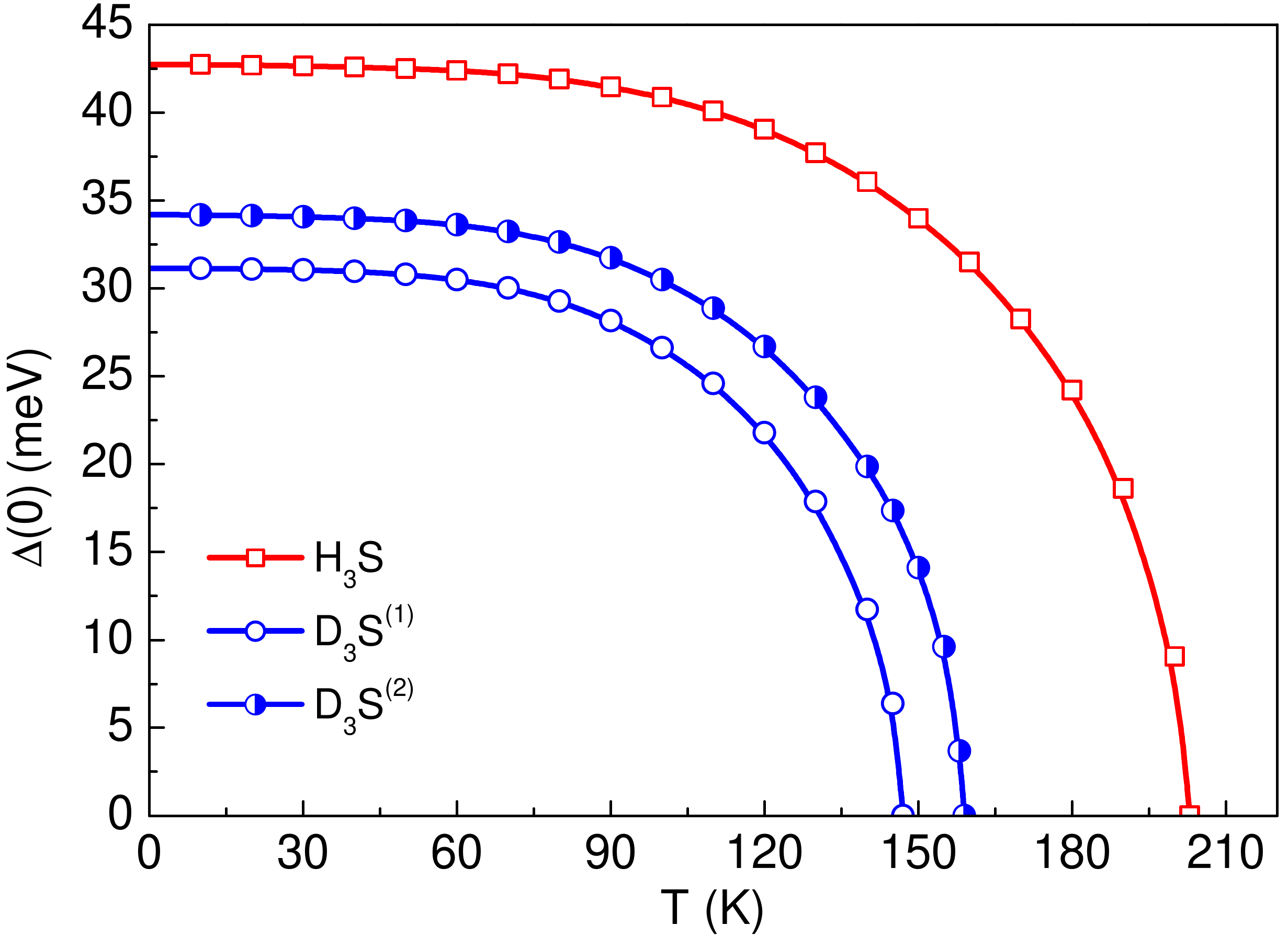}
\caption{The full dependence of the superconducting energy gap on the temperature.}
\label{f3}
\end{figure}
%
\begin{figure}[!b]
\includegraphics[width=\columnwidth]{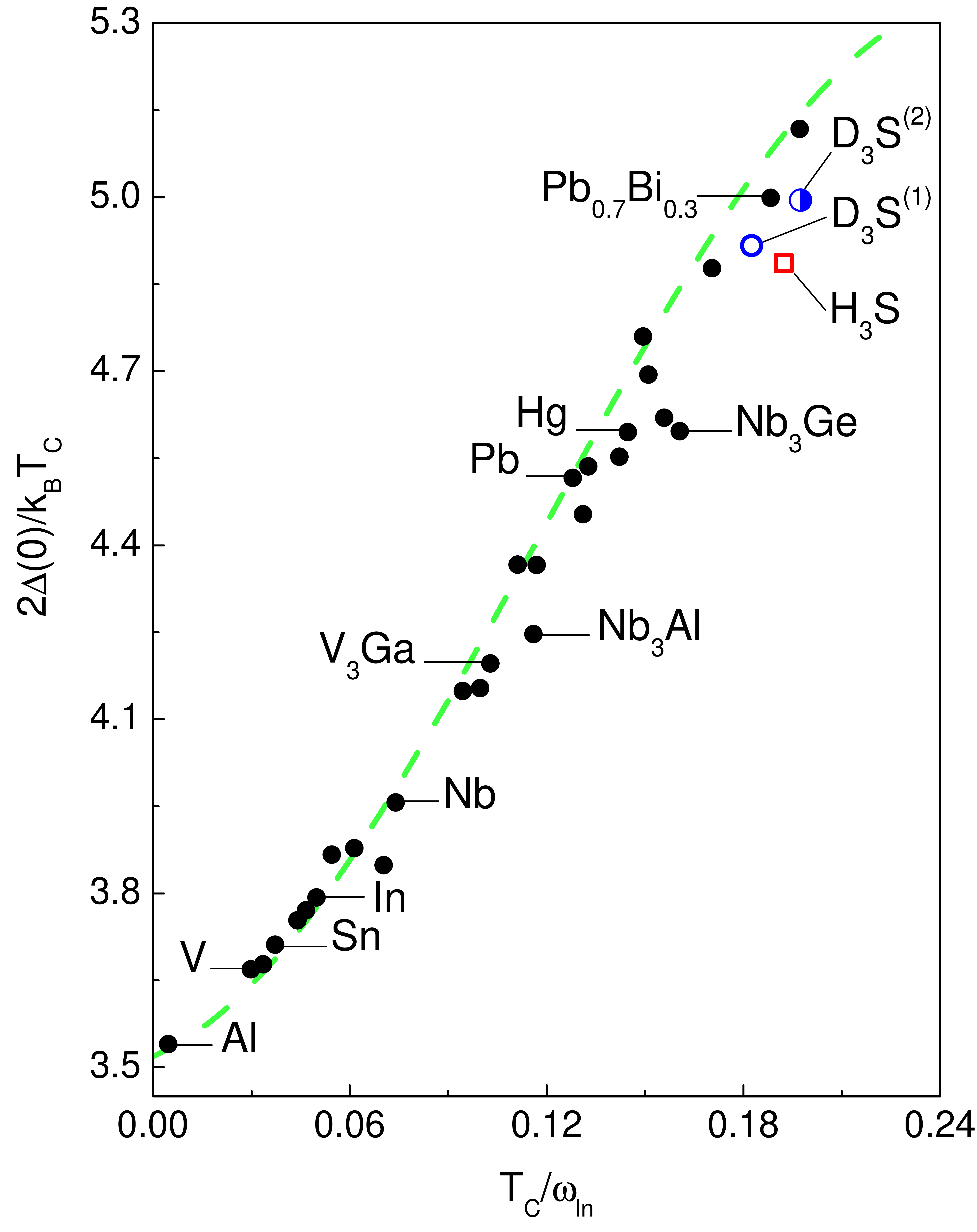}
\caption{The superconducting gap ratio as a function of $T_C/\omega_{ln}$. The results for the selected conventional superconductors are taken from paper \cite{Carbotte1990A}. The dashed line are obtained on the basic of paper \cite{Mitrovic}.}
\label{f4}
\end{figure}
%
By using the Eliashberg spectral functions and the determined above Coulomb pseudopotentials as an input parameters to the Eliashberg equations, we obtained the superconducting energy gap values $\Delta(0)=42.74$ meV for $\rm H_3S$ and $\Delta(0)^{\rm (1)}=31.14$ meV and $\Delta(0)^{\rm (2)}=34.21$ meV for $\rm D_3S$, which corresponds to $[2\Delta(0)/k_{B}T_{C}]_{\rm H_3S}=4.89$, $[2\Delta(0)/k_{B}T_{C}]^{\rm (1)}_{\rm D_3S}=4.92$ and $[2\Delta(0)/k_{B}T_{C}]^{\rm (2)}_{\rm D_3S}=4.99$. 
The temperature dependences of the superconducting gaps are shown in \fig{f3}. 
Moreover, we compared the calculated gap ratio for $\rm H_3S$ and $\rm D_3S$ with other conventional superconductors. In particular, in \fig{f4} we can see $2\Delta(0)/k_{B}T_{C}$ as a function of the strong-coupling index $T_C/\omega_{ln}$ \cite{Carbotte2}. 
At this point it should be noted that in the weak coupling BCS limit $T_{C}/\omega_{ln}\rightarrow 0$.
In \fig{f4}, the full circles representing results for selected conventional superconductors and dashed line are taken from papers \cite{Carbotte1990A, Mitrovic}.
It is clearly visible that although the obtained results differ significantly from the prediction of the classic BCS theory ($[2\Delta(0)/k_{B}T_{C}]_{\rm BCS}=3.53$), they are close to the trend determined by the conventional superconductors. The recent study reported by Nicol and Carbotte \cite{Nicol} generally confirmed our calculations.

In the next step, we calculated the specific heat difference between the superconducting and the normal states:
\begin{equation}
\label{rC}
\frac{\Delta C}{k_{B}N(0)}=-\frac{1}{\beta}\frac{d^{2}\left[\Delta F/N(0)\right]}{d\left(k_{B}T\right)^{2}},
\end{equation}
where symbol $\Delta F$ denotes the free energy difference between the superconducting and the normal state  \cite{BardeenStephen}: 
\begin{eqnarray}
\label{rF}
\frac{\Delta F}{N(0)}&=&-\frac{2\pi}{\beta}\sum_{n=1}^{M}
\left(\sqrt{\omega^{2}_{n}+\Delta^{2}_{n}}- \left|\omega_{n}\right|\right)\\ \nonumber
&\times&\left(Z^{S}_{n}-Z^{N}_{n}\frac{\left|\omega_{n}\right|}
{\sqrt{\omega^{2}_{n}+\Delta^{2}_{n}}}\right).
\end{eqnarray}  
\begin{figure}[!b]
\includegraphics[width=\columnwidth]{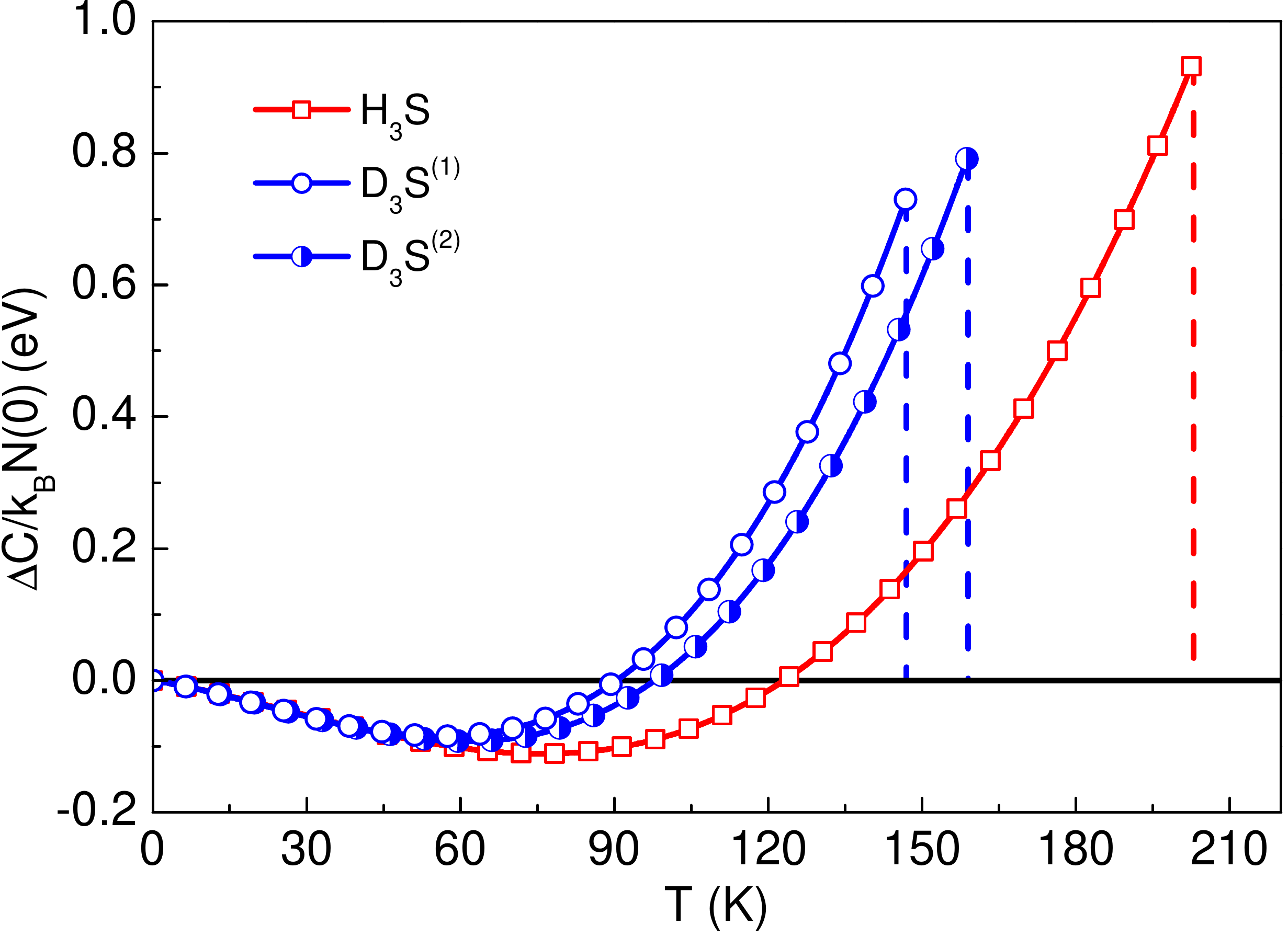}
\caption{The specific heat difference as a function of temperature for H$_3$S and D$_3$S at $150$ GPa.}
\label{f5}
\end{figure}
%
Here, $Z^{S}_{n}$ and $Z^{N}_{n}$ are the mass renormalization functions for the superconducting and for the normal state, respectively.
Let us note that the Eq.\eq{rF} can be easily computed, once the imaginary axis Eliashberg equations \eq{r4} and \eq{r5} are solved.

The specific heat difference for H$_3$S and D$_3$S as a function of temperature was shown in \fig{f5}. The characteristic specific heat jump at $T_C$ was marked by vertical dashed lines.
At critical temperature, the BCS result for the dimensionless ratio $\Delta C\left(T_{C}\right)/{C^{N}\left(T_{C}\right)}$ is equal to $1.43$, and is universal, like the superconducting gap ratio. In the above ratio, the specific heat for the normal state is defined as: $C^{N}=\gamma T$, where symbol $\gamma$ denotes the Sommerfeld constant: $\gamma\equiv ({2}/{3})\pi^{2}\left(1+\lambda\right)k_{B}^2N(0)$.
The comparison of the specific heat ratio as a function of $T_C/\omega_{ln}$ for the investigated systems with other conventional superconductors looks similar to the gap ratio (see \fig{f6}). However, despite the fact that the value calculated for D$_3$S at $150$ GPa is very close to the dashed line adopted from paper \cite{MarsiglioCarbotte}, the value for H$_3$S clearly departs from this trend.
\begin{figure}[!t]
\includegraphics[width=\columnwidth]{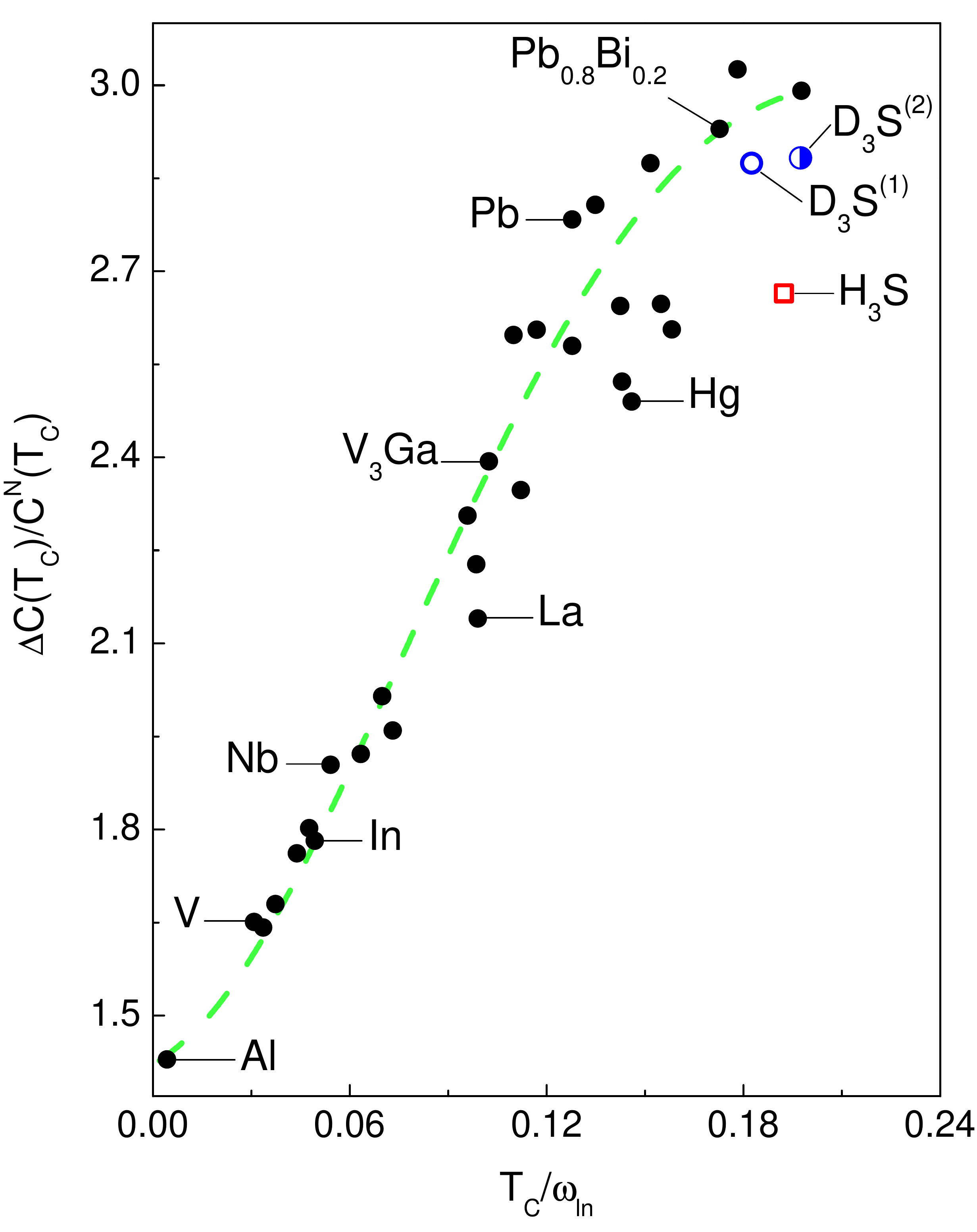}
\caption{The specific heat ratio as a function of $T_C/\omega_{ln}$. The results for the selected conventional superconductors are taken from paper \cite{Carbotte1990A}. The dashed line are obtained on the basic of paper \cite{MarsiglioCarbotte}. For H$_{3}$S, D$_{3}$S$^{(1)}$ and D$_{3}$S$^{(2)}$ the ratio $\Delta C\left(T_{C}\right)/{C^{N}\left(T_{C}\right)}$ is equal to $2.66$, $2.87$ and $2.88$, respectively.}
\label{f6}
\end{figure}
%

In the next step, on the basis of Eq.\eq{rF}, the thermodynamic critical field $H_{C}$ and the deviation function of the thermodynamic critical field $D$ can be obtained from the following relations:

%
\begin{equation}
\label{rH}
\frac{H_{C}}{\sqrt{N(0)}}=\sqrt{-8\pi\left[\Delta F/N(0)\right]}
\end{equation}
and
\begin{equation}
D = {H_c\left(T\right)\over{H_c\left(0\right)}}-\left[1-\left({T\over{T_c}}\right)^2\right].
\label{rD}
\end{equation}
%
The influence of the temperature on the thermodynamic critical field was presented in \fig{f7} (A).
In \fig{f7} (B) we supplement our results with the calculated thermodynamic critical field deviation as a function of the reduced temperature. The results predicted by the weak-coupling BCS theory are presented with a dashed line \cite{Michor}.
As for the investigated materials our results confirm the non-BCS behaviour of the thermodynamic critical field deviation. 
\begin{figure}[!b]
\includegraphics[width=\columnwidth]{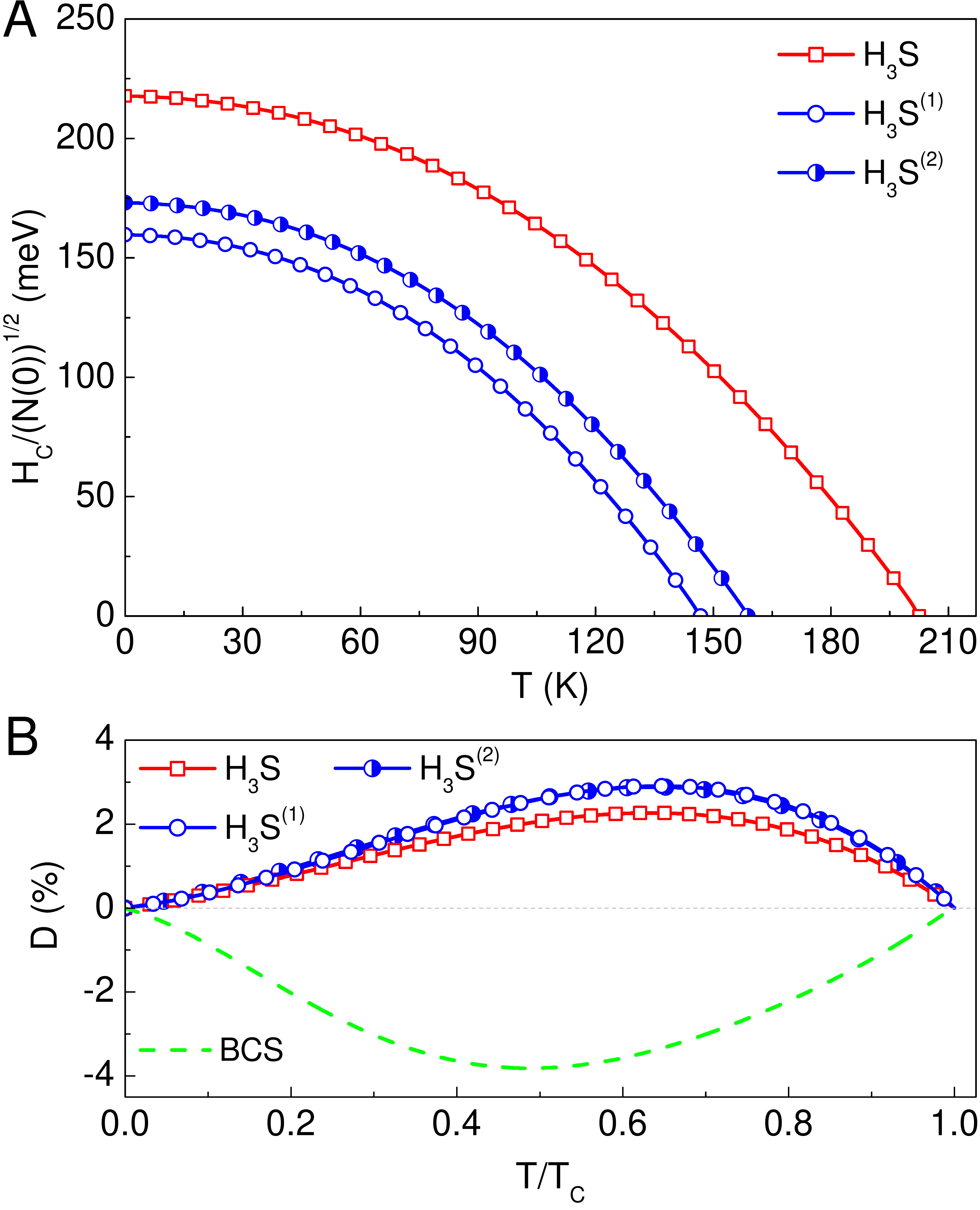}
\caption{(A) The thermodynamic critical field and (B) the critical field deviation as a function of temperature for H$_3$S and D$_3$S at $150$ GPa.}
\label{f7}
\end{figure}
In the BCS theory, the dimensionless ratio connected with thermodynamic critical field ${T_{C}C^{N}\left(T_{C}\right)}/{H_{C}^{2}\left(0\right)}$ also takes the universal value equal to $0.168$.
In the case of hydrogen and deuterium sulfide we observed a high derogation from the BCS theory prediction, in particular 
$[{T_{C}C^{N}\left(T_{C}\right)}/{H_{C}^{2}\left(0\right)}]_{\rm H_3S}=0.130$, $[{T_{C}C^{N}\left(T_{C}\right)}/{H_{C}^{2}\left(0\right)}]^{\rm (1)}_{\rm D_3S}=0.127$ and $[{T_{C}C^{N}\left(T_{C}\right)}/{H_{C}^{2}\left(0\right)}]^{\rm (2)}_{\rm D_3S}=0.126$. The comparison with other superconductors is presented in \fig{f8}.
Also here the result computed for $\rm H_3S$ clearly differs from the general trend set by the dashed line \cite{MarsiglioCarbotte}.

\begin{figure}[!t]
\includegraphics[width=\columnwidth]{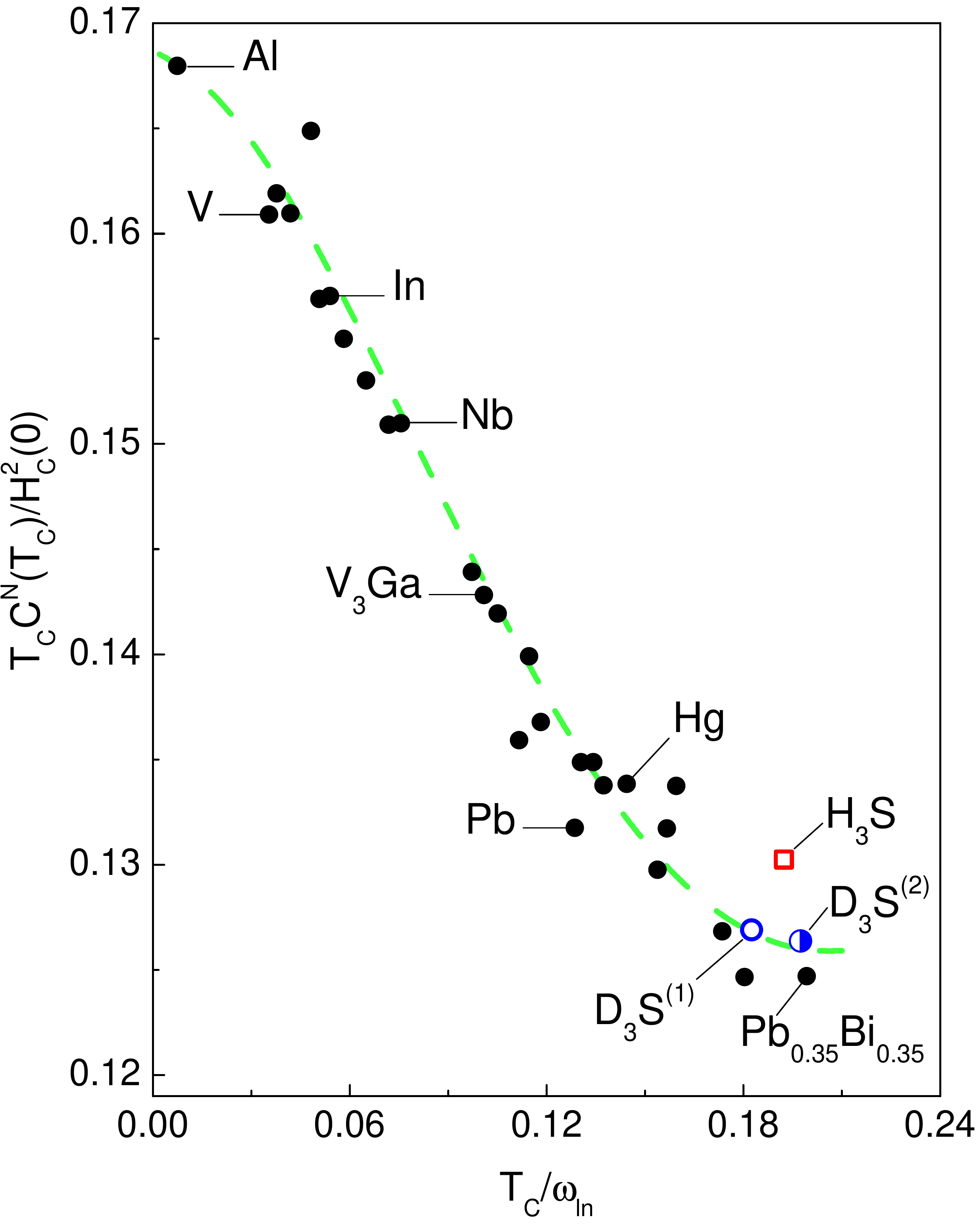}
\caption{The thermodynamic critical field ratio as a function of $T_C/\omega_{ln}$. The results for the selected conventional superconductors are taken from paper \cite{Carbotte1990A}. The dashed line are obtained on the basic of paper \cite{MarsiglioCarbotte}.}
\label{f8}
\end{figure}
%
\newpage
In the last step, we computed the London penetration depth ($\lambda_{L}$):
\begin{eqnarray}
\label{rLPD}
\frac{1}{e^2v^2_F N(0)\lambda^2_L\left(T\right)}=\frac{4}{3}\frac{\pi}{\beta}\sum_{n=1}^{M}
\frac{\Delta^{2}_{n}}{Z^{S}_{n}\left[ \omega^{2}_{n}+\Delta^{2}_{n}\right]^{3/2}},
\end{eqnarray}
%
\begin{figure}[!b]
\includegraphics[width=\columnwidth]{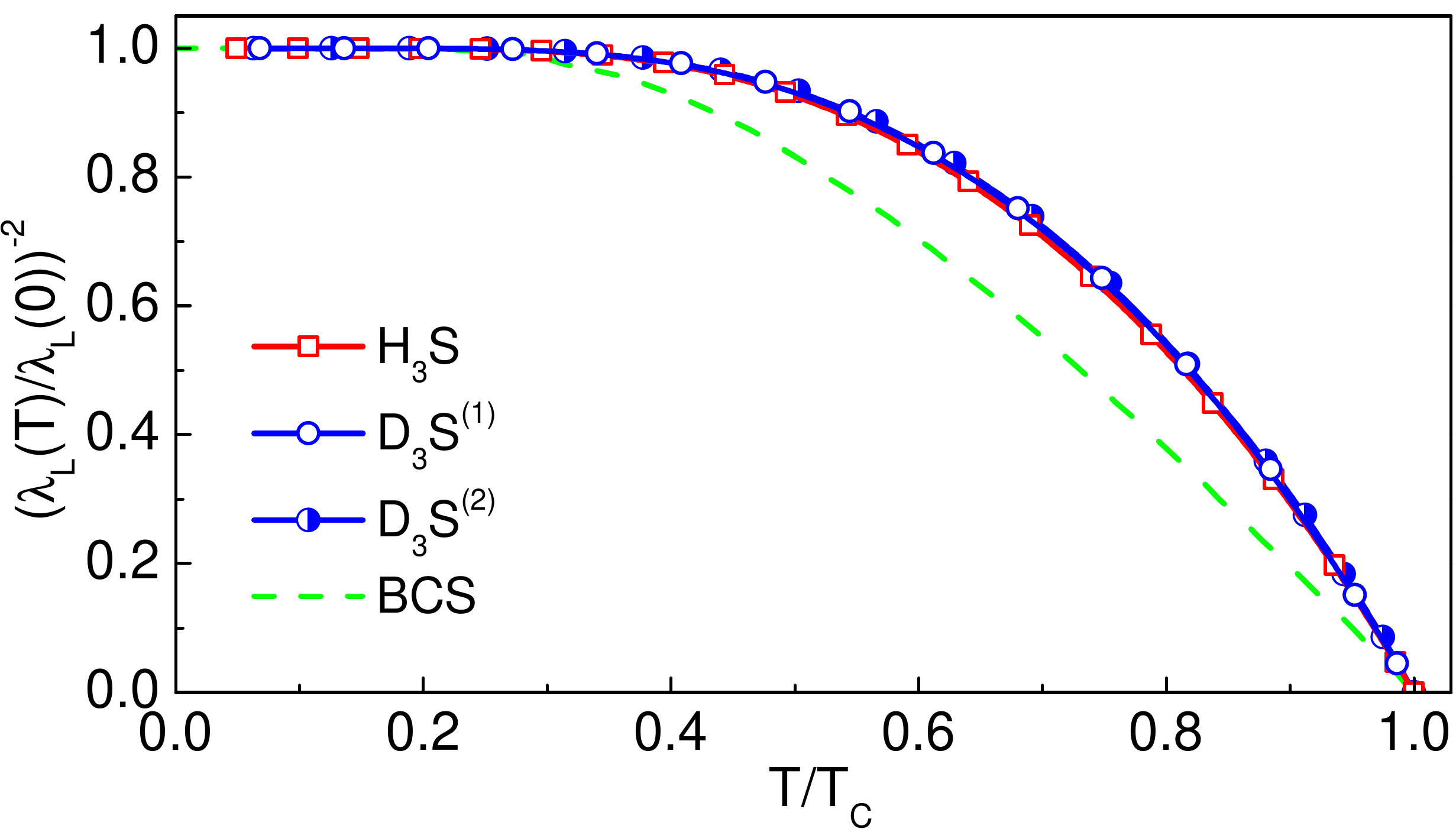}
\caption{The normalized London penetration depth as a function of temperature.}
\label{f9}
\end{figure}
where $e$ is the electron charge and $v^{}_F$ is the Fermi velocity \cite{Carbotte1990A}.
The normalized $\lambda^{-2}_L$ as a function of temperature is presented and compared with the BCS theory prediction in \fig{f9}.

The differences observed between the BCS and the Eliashberg theory are connected with the fact that the BCS approach does not include the strong coupling corrections and the retardation effect of the actual electron-phonon interactions as described by the Eliashberg theory.
\\\\

\section{IV. Conclusions}
%
In this paper, using the phonon-mediated Eliashberg theory, we have investigated the thermodynamic properties of superconducting hydrogen and deuterium sulfide at high pressure ($p=150$ GPa).
In particular, the energy gap, the specific heat, the thermodynamic critical field and the London penetration depth 
were calculated for Coulomb pseudopotentials $[\mu^{\star}_C]_{\rm H_3S}=0.123$ and $[\mu^{\star}_C]^{\rm (1)}_{\rm D_3S}=0.131$ determined on the basis of experimental values of critical temperature ($[T_C]_{\rm H_3S}=203$ K, $[T_C]^{\rm (1)}_{\rm D_3S}=147$ K) and $[\mu^{\star}_C]^{\rm (2)}_{\rm D_3S}=0.088$ determined on the basis of our prediction of critical temperature for deuterium sulfide ($[T_C]^{\rm (2)}_{\rm D_3S}=159$ K).
Next, on the basis of the calculated thermodynamic functions, it was proven that the values of the dimensionless ratios 
$2\Delta(0)/k_{B}T_{C}$, $\Delta C\left(T_{C}\right)/{C^{N}\left(T_{C}\right)}$ and $T_{C}C^{N}\left(T_{C}\right)/H_{C}^{2}\left(0\right)$ 
differ significantly from the predictions of the BCS theory.
These discrepancies arise from the existence of the strong-coupling and retardation effects in the systems investigated. The Eliashberg theory goes beyond the BCS theory to include these effects.

The results presented in this paper, are expected to stimulate experimental and theoretical exploration and discovery of new superconducting hydrogen-containing materials like H$_3$S. 
Moreover we hope that future measurements will be able to determine conclusively the pressure dependence of critical temperature for hydrogen/deuterium sulfide.

\section{V. Acknowledgments}
%
Artur Durajski gratefully acknowledges financial support from Prof. Z. Stradomski, Prof. Z. Nitkiewicz and from the Cz{\c{e}}stochowa University of Technology under Grant No. BS/MN-203-303/2015.\\
Moreover, the authors would like to thank E. Cappelluti, L. Ortenzi and G. Chiarotti (University of Rome La Sapienza) for scientific discussion, and Ryosuke Akashi (University of Tokyo) for sharing the Eliashberg functions.
%
%
%
\bibliography{H3S}
%
%
\end{document}